\documentclass[11pt]{article}
\usepackage{amssymb}
\usepackage{amsmath}
\usepackage{graphics}
\usepackage{epsfig}
\usepackage{a4wide}
\usepackage{cite}
\usepackage{multirow}
\usepackage{color}
\usepackage{mathrsfs}
\usepackage{graphicx}
\usepackage{slashed}

\DeclareGraphicsExtensions{.eps}
\begin{document}

\title{Precision study of $W^-W^+H$ production including parton shower effects at the CERN Large Hadron Collider}
\author{
Huan-Yu Bi$^{1,2}$, Ren-You Zhang$^{1,2}$\footnote{Corresponding author. zhangry@ustc.edu.cn}, Wen-Gan Ma$^{1,2}$, Yi Jiang$^{1,2}$, Xiao-Zhou Li$^{1,2}$, \\ Peng-Fei Duan$^{3}$
\\ \\
{\small $^1$ State Key Laboratory of Particle Detection and Electronics,} \\
{\small University of Science and Technology of China, Hefei 230026, Anhui, People's Republic of China} \\
{\small $^2$ Department of Modern Physics, University of Science and Technology of China,}  \\
{\small Hefei 230026, Anhui, People's Republic of China} \\
{\small $^3$ City College, Kunming University of Science and Technology,}  \\
{\small Kunming 650051, Yunnan, People's Republic of China}
}

\date{}
\maketitle
\vskip 10mm

\begin{abstract}
The precision study of $W^-W^+H$ production with subsequent $W^{\pm} \rightarrow l^{\pm} \overset{ _{(-)}}{\nu_{l}}$ and $H \rightarrow b\bar{b}$ decays at the LHC can help us to study the Higgs gauge couplings and to search for new physics beyond the SM. In this paper, we calculate the shower-matched NLO QCD correction and the EW corrections from the $q\bar{q}$ annihilation and photon-induced channels to the $W^-W^+H$ production at the $14~ {\rm TeV}$ LHC, and deal with the subsequent decays of Higgs and $W^{\pm}$ bosons by adopting the {\sc MadSpin} method. Both the integrated cross section and some kinematic distributions of $W^{\pm}$, $H$ and their decay products are provided. We find that the QCD correction enhances the LO differential cross section significantly, while the EW correction from the $q\bar{q}$ annihilation channel obviously suppresses the LO differential cross section, especially in the high energy phase-space region due to the Sudakov effect. The $q\gamma$- and $\gamma\gamma$-induced relative corrections are positive, and insensitive to the transverse momenta of $W^{\pm}$, $H$ and their decay products. These photon-induced corrections compensate the negative $q\bar{q}$-initiated EW correction, and become the dominant EW contribution as the increment of the $pp$ colliding energy. The parton shower (PS) effects on the kinematic distributions are nonnegligible. The PS relative correction to the $b$-jet transverse momentum distribution can exceed $100\%$ in the high $p_{T, b}$ region. We also investigate the scale and PDF uncertainties, and find that the theoretical error of the ${\rm QCD}+{\rm EW}+q\gamma+\gamma\gamma$ corrected integrated cross section mainly comes from the renormalization scale dependence of the QCD correction.
\end{abstract}


\vfill \eject
\baselineskip=0.32in
\makeatletter      
\@addtoreset{equation}{section}
\makeatother       
\vskip 5mm
\renewcommand{\theequation}{\arabic{section}.\arabic{equation}}
\renewcommand{\thesection}{\Roman{section}.}
\newcommand{\nb}{\nonumber}
\section{INTRODUCTION}
\par
The Higgs mechanisim is responsible for the electroweak (EW) symmetry breaking and the origin of masses of elementary particles \cite{Glashow:1961tr,Weinberg:1967tq,Salam:1968rm,Higgs:1964pj,Englert:1964et}, thus plays an important role in the Standard Model (SM). The Higgs boson has been discovered in 2012 at the CERN Large Hadron Collider (LHC) \cite{Aad:2012tfa,Chatrchyan:2012xdj}. One of the main tasks nowadays at the LHC is to detailedly study the spin and CP properties of the Higgs boson, and the Higgs gauge and Yukawa interactions. In order to understand the Higgs boson in the most accurate way, one should study not only the main Higgs production channels, but also the rare processes that can be sensitive to new physics \cite{Gabrielli:2013era}.

\par
The precision phenomenological study of $VV^{\prime}H~ (V,~ V^{\prime} = W~{\rm or}~ Z)$ productions is helpful for the study of Higgs gauge couplings, since it can be used to determine the ratio of $WWH$ coupling to $ZZH$ coupling \cite{Chiang:2018fqf} and to study the Higgs anomalous gauge couplings \cite{Gabrielli:2013era}. The EW corrections to $VV^{\prime}H$ productions at the LHC are directly related to the triple and quartic gauge couplings, such as $WWZ$, $WW\gamma$, $WWZZ$, $WWZ\gamma$, $WW\gamma\gamma$ and $WWWW$ couplings. $pp \rightarrow VV^{\prime}H + X$ processes also contain $HHVV^{\prime}$ couplings. Since there is no constraints on these Higgs quartic gauge couplings so far, detecting the Higgs production in association with two gague bosons at the LHC and the future high energy hadron colliders (FCC-hh and SPPC) may help us to constrain the bounds on the Higgs quartic gauge couplings (even though the backgrounds to these processes could be very large) \cite{Agrawal:2019ffb}.

Strictly speaking, the EW symmetry breaking is determined by the shape of the Higgs potential. However, from another point of view, the Higgs triple gauge couplings are also related to the EW symmetry breaking: The gauge invariance of the Higgs kinetic term implies the existence of the Higgs quartic gauge interactions. These Higgs quartic gauge interactions would induce Higgs triple gauge interactions as well as the masses of the weak gauge bosons once the EW symmetry is spontaneously broken. Since the $VV^{\prime} H$ productions at the LHC are directly related to the Higgs triple gauge couplings, the precision study of $pp \rightarrow VV^{\prime} H + X$ can help to understand the EW symmetry breaking and to search for new physics beyond the SM.

\par
Scrutinizing Higgs properties needs accurate theoretical predictions and precise experimental measurements on both signals and backgrounds. At the LHC, $W^-W^+b\bar{b}$ channel is an important final state. Many SM and BSM processes are measured though this final state, such as top pair production, Higgs pair production \cite{CMS:2015nat,CMS:2016cdj} and vector-like quark pair production\cite{Sirunyan:2017pks}. The production channel $pp \rightarrow W^-W^+H \rightarrow W^-W^+ b\bar{b}$ could also be an irreducible background to these processes. The top pair production with subsequent decay $t \rightarrow W b$ at the LHC has been widely investigated over the past twenty years, and the Higgs pair production and decays into $W^-W^+b\bar{b}$, $\gamma\gamma b\bar{b}$ and $\tau^-\tau^+b\bar{b}$ final states have been studied at the High-Luminosity LHC \cite{CMS:2015nat}. The $W^-W^+H$ production at the QCD next-to-leading order (NLO) including parton shower (PS) matching has been investigated in Refs. \cite{Baglio:2015eon,Baglio:2016ofi,Mao:2009jp}. Further precision study of the $W^-W^+H$ production should involve the shower-matched NLO QCD (QCD+PS) correction, the $q\bar{q}$-, $q\gamma$- and $\gamma\gamma$-initiated EW corrections, and the subsequent decays of $W^{\pm}$ and Higgs bosons.

\par
In this work, we study in detail the $W^-W^+H$ production with subsequent $W^{\pm} \rightarrow l^{\pm} \overset{ _{(-)}}{\nu_{l}}$ and $H \rightarrow b\bar{b}$ decays at the LHC, i.e., $pp \rightarrow W^-W^+H \rightarrow l^+ l^- \nu_{l} \bar{\nu}_{l} b \bar{b} +X~ (l = e~ {\rm or}~ \mu)$, including the QCD+PS correction and the EW corrections from the $q\bar{q}$ annihilation and photon-induced channels. The rest of this paper is organized as follows. In Sec. II we describe in detail the analytical calculation strategy. In Sec. III we present the numerical results of the integrated cross section and some kinematic distributions, and discuss the theoretical uncertainties from the factorization and renormalization scales and parton distribution functions (PDFs). Finally, a short summary is given in Sec. IV.

\section{CALCULATION STRATEGY}
\par
\begin{figure}[htbp]
\begin{center}
\includegraphics[scale=0.6]{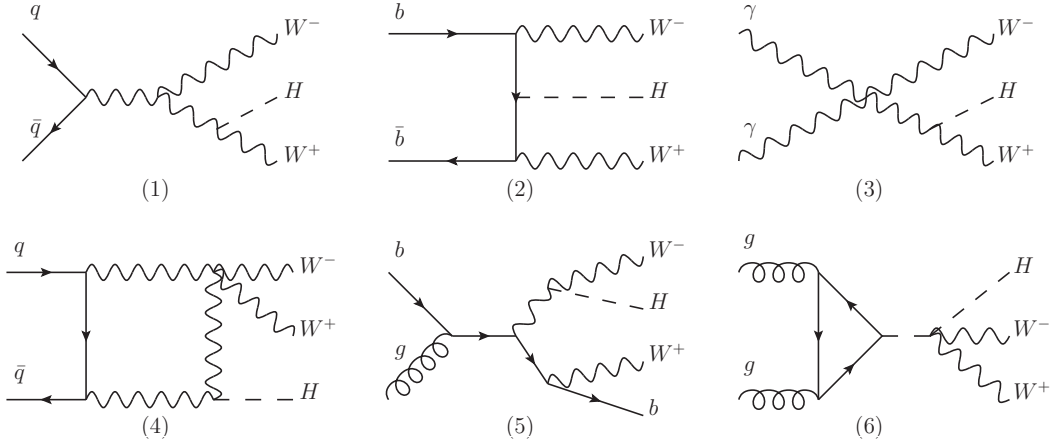}
\caption{Representative Feynman diagrams for the partonic processes contributing to the $pp \rightarrow W^-W^+H+X$ process.}
\label{fig1}
\end{center}
\end{figure}

In this paper, the precision calculation for the $pp \rightarrow W^-W^+H + X$ process involves the following partonic channels: (1) quark-antiquark annihilation $q \bar{q} \rightarrow W^-W^+H + (g/\gamma)$, (2) real light-quark emission $q g/\gamma \rightarrow W^-W^+H + q$, (3) gluon-gluon fusion $gg \rightarrow W^-W^+H$, and (4) photon-photon fusion $\gamma\gamma \rightarrow W^-W^+H$, where $q$ runs over all five light flavors of quarks. The $q\bar{q}$ annihilation subprocesses are calculated up to the QCD+EW NLO,
\begin{eqnarray}
\sigma_{q\bar{q}} = \sigma_{q\bar{q}}^{{\rm LO}} + \Delta \sigma_{q\bar{q}}^{{\rm QCD}} + \Delta \sigma_{q\bar{q}}^{{\rm EW}},
\end{eqnarray}
where $\sigma_{q\bar{q}}^{{\rm LO}}$, $\Delta \sigma_{q\bar{q}}^{{\rm QCD}}$ and $\Delta \sigma_{q\bar{q}}^{{\rm EW}}$ are the ${\cal O}(\alpha^3)$, ${\cal O}(\alpha^3\alpha_s)$ and ${\cal O}(\alpha^4)$ contributions from the $q\bar{q}$ annihilation subprocesses, respectively. The subprocesses with $qg$ and $q\gamma$ initial states are calculated only at the LO, and the corresponding cross sections are denoted as $\sigma_{qg}$ and $\sigma_{q\gamma}$. It should be noted that the PDF counterterm corrections from the $q \rightarrow q + g$, $q \rightarrow q + \gamma$, $g \rightarrow q + \bar{q}$ and $\gamma \rightarrow q + \bar{q}$ parton splittings should be included into $\Delta \sigma_{q\bar{q}}^{{\rm QCD}}$, $\Delta \sigma_{q\bar{q}}^{{\rm EW}}$, $\sigma_{qg}$ and $\sigma_{q\gamma}$, respectively, for IR safety\footnote{There are two types of PDF counterterm corrections from the $q \rightarrow q + \gamma$ splitting, which correspond to the $P_{qq}$ and $P_{\gamma q}$ splitting functions, respectively. The EW PDF counterterm correction induced by the splitting function $P_{qq}$ is absorbed into $\Delta \sigma_{q\bar{q}}^{{\rm EW}}$, while the correction induced by the splitting function $P_{\gamma q}$ as well as the PDF counterterm correction from the $\gamma \rightarrow q + \bar{q}$ splitting is absorbed into $\sigma_{q\gamma}$.}. Due to the large gluon density in proton at high energy hadron colliders, the loop-induced channel $gg \rightarrow W^-W^+H$ is taken into account in our precision QCD calculation, although the LO contribution of the $gg$ fusion channel is one order of $\alpha_s$ higher than the NLO QCD correction from the $q\bar{q}$ annihilation channel. For the $\gamma\gamma$ fusion channel, the NLO EW correction is negligible and we consider only its LO contribution to the $pp \rightarrow W^-W^+H + X$ process, because the density of photon in proton is much less than those of colored partons (i.e., gluon and light quarks). We generate all the Feynman diagrams and amplitudes for these partonic channels by adopting {\sc FeynArts} package \cite{Hahn:2000kx}, and present some representative Feynman diagrams in Fig.\ref{fig1}. Then the corrected cross section for the $pp \rightarrow W^-W^+H + X$ process without the matching to parton shower calculated in this paper is given by
\begin{eqnarray}
\sigma^{{\rm QCD}+{\rm EW}+q\gamma+\gamma\gamma} = \sigma^{{\rm LO}} + \Delta \sigma^{{\rm QCD}} + \Delta \sigma^{{\rm EW}} + \sigma_{q\gamma} + \sigma_{\gamma\gamma},
\end{eqnarray}
where the LO cross section, QCD correction and EW correction are defined as\footnote{In this paper, we define $\Delta \sigma^{{\rm EW}}$ as the EW correction from the $q\bar{q}$ annihilation channel in order to show the photon-induced contributions ($\sigma_{q\gamma}$ and $\sigma_{\gamma\gamma}$) more clearly.}
\begin{eqnarray}
\sigma^{{\rm LO}} = \sigma_{q\bar{q}}^{{\rm LO}},~~~~~~~~
\Delta \sigma^{{\rm QCD}} = \Delta \sigma_{q\bar{q}}^{{\rm QCD}} + \sigma_{qg} + \sigma_{gg},~~~~~~~~
\Delta \sigma^{{\rm EW}} = \Delta \sigma_{q\bar{q}}^{{\rm EW}},
\end{eqnarray}
and $\sigma_{gg} \sim {\cal O}(\alpha^3\alpha_s^2)$ and $\sigma_{\gamma\gamma} \sim {\cal O}(\alpha^3)$ are the lowest order contributions of the $gg$ and $\gamma\gamma$ fusion channels, respectively. Sine the calculation of the NLO QCD correction has been presented in Refs.\cite{Baglio:2015eon,Baglio:2016ofi,Mao:2009jp}, we describe only the calculation of the EW correction in this section.

\par
In the calculation of $\sigma^{{\rm LO}}$, $\Delta \sigma^{{\rm QCD}}$, $\Delta \sigma^{{\rm EW}}$ and $\sigma_{q\gamma}$, we adopt the $G_{\mu}$ scheme \cite{Nhung:2013jta,Sirlin:1980nh,Yong-Bai:2016sal} (i.e., $\alpha = \alpha_{G_{\mu}}$) for all the EW couplings. This fine structure constant scheme is suitable for EW correction due to the large EW Sudakov logarithms caused by the soft or collinear weak gauge-boson exchange at high energies \cite{Denner:2014bna,Denner:2015fca}. But in the evaluation of the $\gamma\gamma$ fusion channel, we adopt the mixed scheme \cite{Andersen:2014efa}, in which the fine structure constant is taken as $\alpha = \alpha(0)$ and $\alpha = \alpha_{G_{\mu}}$ for the electromagnetic and weak couplings, respectively. In the mixed scheme, the mass-singular terms $\ln(m_f^2/\mu^2)~ (f = e, \mu, \tau, u, d, c, s, b)$ from vacuum polarization at the EW NLO can be either canceled between the external photons and the corresponding electromagnetic couplings or absorbed into $\alpha_{G_{\mu}}$ in genuine weak couplings. Thus, the mixed scheme is more suitable for performing high-order perturbative calculation for the processes with external photon legs\footnote{Although the $\gamma\gamma$ fusion channel is calculated only at the LO because the NLO correction is negligible, we still suggest adopting the mixed scheme for this channel.}.

\par
\begin{figure}[htbp]
\begin{center}
\includegraphics[scale=0.35]{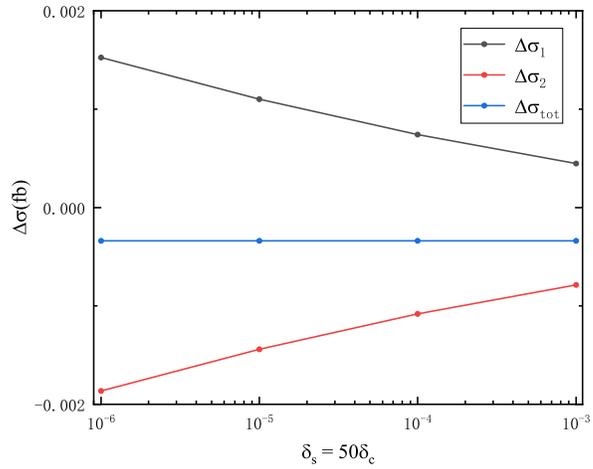}
\caption{$\delta_s$ dependence of the NLO EW corrections to the $W^-W^+H$ production from $u\bar{u}$ annihilation at the LHC. $\Delta \sigma_1$ and $\Delta \sigma_2$ represent the hard-noncollinear correction and soft+collinear+virtual correction, respectively, and $\Delta \sigma_{\rm tot} = \Delta \sigma_1 + \Delta \sigma_2$.}
\label{fig2}
\end{center}
\end{figure}
\begin{figure}[htbp]
\begin{center}
\includegraphics[scale=0.35]{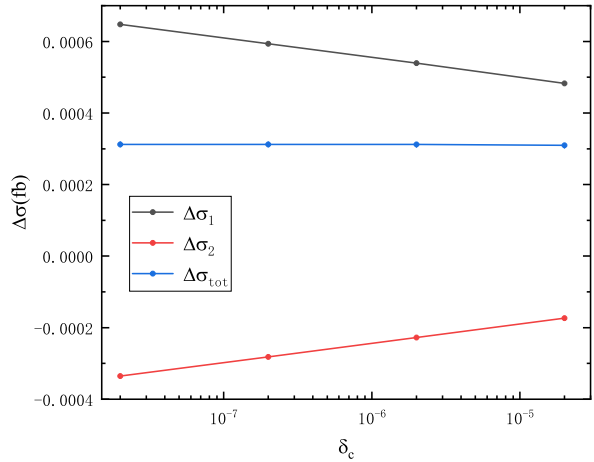}
\caption{$\delta_c$ dependence of the NLO EW corrections to the $W^-W^+H$ production from $u(\bar{u})\gamma$ scattering at the LHC. $\Delta \sigma_1$ and $\Delta \sigma_2$ represent the noncollinear correction and collinear correction, respectively, and $\Delta \sigma_{\rm tot} = \Delta \sigma_1 + \Delta \sigma_2$.}
\label{fig3}
\end{center}
\end{figure}

The ultraviolet (UV) and infrared (IR) divergences appeared in the NLO EW calculation are regularized in dimensional regularization scheme. The complicated 5-point loop integrals encountered in the virtual correction can be decomposed into 4-point loop integrals by employing the Passarino-Veltman algorithm\cite{Passarino:1978jh}. For 4-point loop integrals, the numerical instability induced by small Gram determinant at some phase-space region can be solved by adopting the quadruple precision arithmetic as used in Refs. \cite{Yong-Bai:2016sal,Yu:2014cka}. The electric charge is renormalized in the $G_{\mu}$ scheme, and the relevant fields and masses are renormalized in the on-mass-shell renormalization scheme \cite{Denner:1991kt}. The real emission (i.e., the real photon emission and real light-quark emission) corrections are handled by adopting the two cutoff phase space slicing (TCPSS) method\cite{Harris:2001sx}. In the TCPSS method, two independent cutoff parameters, $\delta_s$ and $\delta_c$, are introduced to decompose the final-state phase space into soft, hard-collinear and hard-noncollinear regions, which are shown schematically in Fig.7 of Ref.\cite{Harris:2001sx}. In this figure, the two triangles marked ``m'' should be included in the hard-collinear region. However, these two triangle regions are excluded since a fixed upper limit of $1 - \delta_s$ is used in calculating the hard-collinear contribution (cf. Eq.(2.35) of Ref.\cite{Harris:2001sx}). We note that the two triangles marked ``m'' give vanishing contribution for $\delta_c \ll \delta_s$. In this work, we set $\delta_c = \delta_s/50$ according to the suggestion of Ref.\cite{Harris:2001sx}. We presented the cutoff dependence of the NLO EW corrections to the $W^-W^+H$ production from $u\bar{u}$ annihilation and $u(\bar{u})\gamma$ scattering at the LHC in Fig.\ref{fig2} and Fig.\ref{fig3}, respectively. From the two figures we may draw the conclusion that the total NLO EW corrections to the $W^-W^+H$ production from different channels are all independent of the cutoff parameters within the calculation errors.
As we expect, the EW correction from the $q\bar{q}$ annihilation channel (i.e., the sum of the virtual correction and the real photon emission correction) and the $q\gamma$-induced correction (i.e., the real light-quark emission EW correction) are both UV and IR finite after absorbing the corresponding PDF counterterms.

\par
The scalar and tensor integrals are calculated by using our developed {\sc LoopTools} package \cite{vanOldenborgh:1990yc}. The PDFs are extracted by LHAPDF6 \cite{Buckley:2014ana}, and the phase-space integration is performed by employing {\sc FormCalc} package \cite{Hahn:1998yk}. The subsequent decays of $W^{\pm}$ and $H$ are handled by using the {\sc MadSpin} method \cite{Artoisenet:2012st}. The matching to parton shower is implemented in the framework of {\sc MadGraph5+Pythia8+MadAnalysis5+FastJet} \cite{Alwall:2014hca,Sjostrand:2014zea,Conte:2014zja,Cacciari:2011ma}. In order to verify the correctness of our calculation, we recalculate the NLO QCD correction to $W^-W^+H$ production with the same input parameters as in Refs. \cite{Baglio:2015eon,Baglio:2016ofi}, and find that our numerical results are in good agreement with the corresponding ones in Refs. \cite{Baglio:2015eon,Baglio:2016ofi} within the calculation errors. We also calculate the NLO EW correction to $ZZH$ production by using our program and obtain $\delta_{{\rm EW}} \simeq 9\%$, which is consistent with that obtained by using the newly developed {\sc MadGraph} package \cite{Frederix:2018nkq}.

\par
A great number of $W^-W^+H$ events are from the $W^-tH$ and $W^+\bar{t}H$ associated productions with subsequent top-quark decay $t \rightarrow Wb$, i.e., $pp \rightarrow bg/\gamma \rightarrow W^-tH \rightarrow W^-W^+Hb + X$ and $pp \rightarrow \bar{b}g/\gamma \rightarrow W^+\bar{t}H \rightarrow W^-W^+H\bar{b} + X$ (see Fig.\ref{fig1}(5)). These events should be treated as the single top production, and thus should be subtracted carefully from our calculation to avoid double counting and to keep the convergence of the perturbative description of the $W^-W^+H$ production. In this paper, we introduce four schemes to subtract the on-shell $W^-tH$ and $W^+\bar{t}H$ events in handling the $bg/\gamma$- and $\bar{b}g/\gamma$-induced subprocesses. In scheme I, we assume the event with a final $b$-jet can be rejected with $100\%$ efficiency, so that the $pp \rightarrow b g/\gamma \rightarrow W^-W^+Hb+X$ and $pp \rightarrow \bar{b} g/\gamma \rightarrow W^-W^+H\bar{b}+X$ event samples can be easily excluded \cite{Baglio:2016ofi}. In scheme II, we adopt the diagram subtraction (DS) method \cite{Liang-Wen:2014fla,Zhang:2011qz} to subtract the top-resonance effect. This subtraction scheme is defined as a replacement of the Breit-Wigner propagator
\begin{eqnarray}
\frac{\vert {\cal M} \vert^2(p_t^2)}{(p_t^2-m_t^2)^2 + \Gamma_t^2 m_t^2}
\rightarrow
\frac{\vert {\cal M} \vert^2(p_t^2)}{(p_t^2-m_t^2)^2 + \Gamma_t^2 m_t^2}
-
\frac{\vert {\cal M} \vert^2(m_t^2)}{(p_t^2-m_t^2)^2 + \Gamma_t^2 m_t^2}
\Theta\big(\sqrt{\hat{s}} - M_W - m_t - M_H \big),~~
\end{eqnarray}
where $p_t^2$ is the squared momentum flowing through the intermediate top-quark propagator and $\sqrt{\hat{s}}$ represents the parton-level colliding energy. In this scheme, the contributions from the squared amplitudes with on-shell top quark are removed point by point over the entire phase space, and the gauge invariance is guaranteed in the limit $\Gamma_t \rightarrow 0$. In scheme III, we adopt the diagram removal (DR) method \cite{Frixione:2008yi,Hollik:2012rc}, i.e., remove all the top-resonance diagrams at the amplitude level, to subtract the top-resonance effect. This DR method violates gauge invariance. However, the authors in Refs. \cite{Frixione:2008yi,Hollik:2012rc} investigated in detail the gauge dependence for the $Wt$ and squark-pair productions and found that the influence of gauge dependence in the DR scheme can be safely neglected in numerical studies. In scheme IV, we introduce the following subtraction term to remove the contributions from the $W^-tH$ and $W^+\bar{t}H$ productions with subsequent top-quark decay at the cross section level\footnote{We assume ${\rm Br}(t\rightarrow Wb)=100\%$ for simplicity.} \cite{Tait:1999cf},
\begin{eqnarray}
\label{sub-term}
\sigma_{{\rm sub}}
=
-
\Big[
\sigma^{{\rm LO}}(pp \rightarrow bg/\gamma \rightarrow W^-tH + X)
+
\sigma^{{\rm LO}}(pp \rightarrow \bar{b}g/\gamma \rightarrow W^+\bar{t}H + X)
\Big]
\times {\rm Br}(t \rightarrow W b).~~
\end{eqnarray}
This scheme can keep gauge invariance, since there is no diagram removal at the amplitude level. In Refs. \cite{Frixione:2008yi,Belyaev:1998dn} the authors suggest imposing an invariant mass cut on the $W^+b$ and $W^-\bar{b}$ systems, which can be written in the form
\begin{eqnarray}
\label{inv-cut}
\vert M_{Wb} - m_t \vert > \kappa \Gamma_t,
\end{eqnarray}
to exclude the $W^-tH$ and $W^+\bar{t}H$ events, respectively. However, we do not adopt this scheme in our calculation, because the $pp \rightarrow bg/\gamma \rightarrow W^-W^+Hb + X$ and $pp \rightarrow \bar{b}g/\gamma \rightarrow W^-W^+H\bar{b} + X$ subprocesses can not be properly handled by using the TCPSS method after applying the invariant mass cut in Eq.(\ref{inv-cut}).

\section{NUMERICAL RESULTS}
\subsection{Input parameters}
\par
The Fermi constant and mass parameters are taken from the recent CERN Yellow Report ``Handbook of LHC Higgs cross sections: 4. Deciphering the nature of the Higgs sector" \cite{deFlorian:2016spz}:
\begin{eqnarray}
&&
M_W = 80.385~ {\rm GeV},~~~~~
M_Z = 91.1876~ {\rm GeV},~~~~~
m_t = 172.5~ {\rm GeV},\nonumber \\
&&
M_H=125~{\rm GeV},~~~~
G_{\mu} = 1.1663787 \times 10^{-5}~ {\rm GeV}^{-2}.
\end{eqnarray}
The top-quark decay width $\Gamma_t=1.41$ GeV and the fine structure constant in the $\alpha(0)$ scheme $\alpha(0)=1/137.035999139$ are taken from Ref. \cite{Patrignani:2016xqp}. In the ${G_\mu}$ scheme we obtain
\begin{eqnarray}
\alpha = \alpha_{G_\mu}= \frac{\sqrt{2}}{\pi} G_{\mu} M_W^2 \left( 1 - \frac{M_W^2}{M_Z^2} \right).
\end{eqnarray}
The strong coupling constant $\alpha_s$ is taken from the PDFs. The factorization and renormalization scales are set to be equal, i.e., $\mu_f = \mu_r = \mu$, and the central scale is chosen as $\mu_0 = M_T/2$ unless stated otherwise, where $M_T$ is the sum of the transverse masses of final particles. We adopt the ${\rm LUXqed\_plus\_PDF4LHC15\_nnlo\_100}$ PDFs \cite{Manohar:2016nzj} throughout the LO and NLO calculations as used in Refs. \cite{Biedermann:2017oae,Denner:2016wet}. All leptons and quarks except the top quark are treated as massless particles\footnote{In this paper, the bottom-quark mass is set to zero in the calculation of the $pp \rightarrow W^-W^+H + X$ production process, but is kept to be nonzero when considering its subsequent Higgs-boson decay $H \rightarrow b \bar{b}$.}, and the Cabibbo-Kobayashi-Maskawa (CKM) matrix is set to $\mathbf{1}_{3 \times 3}$. The $W$-boson decay branching ratio ${\rm Br}(W^{\pm} \rightarrow l^{\pm} \overset{ _{(-)}}{\nu_{l}}) = 22.2\%$ is obtained by using the {\sc MadSpin} program, and the Higgs-boson decay branching ratio ${\rm Br}(H \rightarrow b \bar{b}) = 57.5\%$ is taken from Ref. \cite{Khachatryan:2016vau}.

\subsection{Integrated cross sections}
\par
In Table \ref{total-cs}, we present the LO and ${\rm QCD}+{\rm EW}+q\gamma+\gamma\gamma$ corrected integrated cross sections and the QCD, EW, $q\gamma$-induced and $\gamma\gamma$-induced corrections ($\Delta\sigma^{{\rm QCD}}$, $\Delta\sigma^{{\rm EW}}$, $\sigma_{q\gamma}$ and $\sigma_{\gamma\gamma}$) for the $W^-W^+H$ production at the $14~ {\rm TeV}$ LHC by employing the four different subtraction schemes mentioned above. The corresponding relative corrections are defined as
\begin{eqnarray}
\delta_{{\rm QCD}} = \dfrac{\Delta\sigma^{{\rm QCD}}}{\sigma^{{\rm LO}}},~~~~~~~~
\delta_{{\rm EW}} = \dfrac{\Delta\sigma^{{\rm EW}}}{\sigma^{{\rm LO}}},~~~~~~~~
\delta_{q\gamma} = \dfrac{\sigma_{q\gamma}}{\sigma^{{\rm LO}}},~~~~~~~~
\delta_{\gamma\gamma} = \dfrac{\sigma_{\gamma\gamma}}{\sigma^{{\rm LO}}}.
\end{eqnarray}
As we expect, $\sigma^{{\rm LO}}$, $\Delta\sigma^{{\rm EW}}$ and $\sigma_{\gamma\gamma}$ are independent of the subtraction scheme, because the subtraction of the $W^-tH$ and $W^+\bar{t}H$ events reduces only the contributions of the $bg/\gamma$ and $\bar{b}g/\gamma$ scattering channels, respectively. The $q\gamma$-induced correction, which is insensitive to the subtraction scheme, is significant ($\Delta\sigma_{q\gamma} \simeq 0.59~ {\rm fb}$, $\delta_{q\gamma} \simeq 6.1\%$), and compensates the negative EW correction from the $q\bar{q}$ annihilation channel ($\Delta\sigma^{{\rm EW}} = -0.58~ {\rm fb}$, $\delta_{{\rm EW}} = -6.0\%$). The contribution from the $\gamma\gamma$ fusion channel is sizable ($\sigma_{\gamma\gamma} = 0.28~ {\rm fb}$, $\delta_{\gamma\gamma} = 2.9\%$), and thus should be taken into account in precision EW calculation, especially when $\delta_{{\rm EW}} + \delta_{q\gamma} \sim 0$. Then the full EW relative correction, defined as $\delta_{{\rm EW}}^{({\rm full})} = \delta_{{\rm EW}} + \delta_{q\gamma} + \delta_{\gamma\gamma}$, is obtained as $\delta_{{\rm EW}}^{({\rm full})} = 2.7\%$ by adopting scheme I. The QCD corrections in scheme I, II and III are almost the same ($\delta_{{\rm QCD}} = 30 \sim 31\%$), while the QCD correction in scheme IV is obviously overestimated since we adopt the narrow-width approximation to subtract the $W^-tH$ and $W^+\bar{t}H$ events in scheme IV (see Eq.(\ref{sub-term})). Since the difference between scheme I, II and III are tiny and the $b$-jet veto can be easily implemented, we adopt only scheme I to deal with the $pp \rightarrow bg/\gamma \rightarrow W^-W^+Hb +X$ and $pp \rightarrow \bar{b}g/\gamma \rightarrow W^-W^+H\bar{b} +X$ subprocesses in the following discussion.
\begin{table}[htbp]
\begin{center}
\renewcommand\arraystretch{1.8}
\begin{tabular}{ccccccc}
\hline
\hline
Subtraction scheme & $\sigma^{{\rm LO}}$ & $\Delta\sigma^{{\rm EW}}$ & $\sigma_{\gamma\gamma}$ & $\sigma_{q\gamma}$ & $\Delta\sigma^{{\rm QCD}}$ & $\sigma^{{\rm QCD}+{\rm EW}+q\gamma+\gamma\gamma}$ \\
\hline
I   & \multirow{4}*{$9.65$} & \multirow{4}*{$-0.58$}~ & \multirow{4}*{$0.28$} & $0.56$ & $2.99$ & $12.90$ \\
II  &                    &                          &                     & $0.59$ & $2.94$ & $12.88$ \\
III &                    &                          &                     & $0.59$ & $2.95$ & $12.89$ \\
IV  &                    &                          &                     & $0.59$ & $3.83$ & $13.77$ \\
\hline
\hline
\end{tabular}
\caption{LO and ${\rm QCD}+{\rm EW}+q\gamma+\gamma\gamma$ corrected integrated cross sections (in fb) for the $W^-W^+H$ production at the $14~ {\rm TeV}$ LHC in subtraction scheme I, II, III and IV.}
\label{total-cs}
\end{center}
\end{table}

\par
The factorization/renormalization scale dependence of the LO and ${\rm QCD}+{\rm EW}+q\gamma+\gamma\gamma$ corrected integrated cross sections and the corresponding QCD, EW, $q\gamma$-induced and $\gamma\gamma$-induced corrections for the $W^-W^+H$ production at the $14~ {\rm TeV}$ LHC are shown in Table \ref{total-scale}. To estimate the theoretical error from the factorization/renormalization scale, we define the scale uncertainty at a given scale $\mu_0$ as\footnote{The scale uncertainties of $\Delta\sigma^{{\rm QCD}}$, $\Delta\sigma^{{\rm EW}}$, $\sigma_{q\gamma}$ and $\sigma_{\gamma\gamma}$ listed in Table \ref{total-scale} are normalized by $\sigma^{{\rm QCD}+{\rm EW}+q\gamma+\gamma\gamma}$, since $\Delta\sigma^{{\rm QCD}}$, $\Delta\sigma^{{\rm EW}}$, $\sigma_{q\gamma}$ and $\sigma_{\gamma\gamma}$ are regarded as the correction components of the corrected cross section $\sigma^{{\rm QCD}+{\rm EW}+q\gamma+\gamma\gamma}$.}
\begin{eqnarray}
\varepsilon_{{\rm scale}}\big(\mu_0\big)
=
\dfrac{1}{\sigma\big(\mu_0\big)}
\max
\left\{
\sigma(\mu) - \sigma(\mu^{\prime}) \,\Big\vert\, \mu,\, \mu^{\prime} \in \big[ \mu_0/2,\, 2 \mu_0 \big]
\right\}.
\end{eqnarray}
We adopt two typical central scales for comparison: (1) $\mu_0^{(1)} = M_T/2$ and (2) $\mu_0^{(2)} = M_F/2$ ($M_F = 2 M_W + M_H$), which are dependent and independent of the final-state phase space, respectively. The scale uncertainties at these two scales are denoted as $\varepsilon_{{\rm scale}}^{(1)}$ and $\varepsilon_{{\rm scale}}^{(2)}$. From Table \ref{total-scale} we can see that whatever the central scale we use, the scale uncertainties of $\sigma^{{\rm LO}}$, $\Delta\sigma^{{\rm EW}}$, $\sigma_{q\gamma}$ and $\sigma_{\gamma\gamma}$ are only about $0.4 \sim 0.6\%$, while the scale uncertainties of $\Delta\sigma^{{\rm QCD}}$ and $\sigma^{{\rm QCD}+{\rm EW}+q\gamma+\gamma\gamma}$ are much significant and exceed $4\%$. It implies that the theoretical error induced by the renormalization scale is roughly one order of magnitude larger than that induced by the factorization scale. We may conclude that the scale uncertainty of the ${\rm QCD}+{\rm EW}+q\gamma+\gamma\gamma$ corrected cross section mainly comes from the renormalization scale dependence of the QCD correction, and the scale uncertainty of the LO cross section is underestimated since the LO cross section does not depend on the strong coupling. The table also shows that the difference between the corrected cross sections at the dynamical scale $\mu_0^{(1)}$ and the fixed scale $\mu_0^{(2)}$ is very small ($\sim 0.7\%$). The EW correction from the $q\bar{q}$ annihilation channel is almost compensated by the $q\gamma$-induced correction, and thus $\delta_{{\rm EW}}^{({\rm full})} \approx \delta_{\gamma\gamma}$, at the $14~ {\rm TeV}$ LHC.
\begin{table}[htbp]
\begin{center}
\renewcommand\arraystretch{1.8}
\begin{tabular}{lcccccccc}
\hline
\hline
\multirow{2}*{}
& \multicolumn{3}{c}{Cross section~(fb)} & ~\multirow{2}*{$\varepsilon_{{\rm scale}}^{(1)}~(\%)$}~
& \multicolumn{3}{c}{Cross section~(fb)} & ~\multirow{2}*{$\varepsilon_{{\rm scale}}^{(2)}~(\%)$}~ \\
\cline{2-4}\cline{6-8}
& $\mu_0^{(1)}/2$ & $\mu_0^{(1)}$ & $2 \mu_0^{(1)}$ &
& $\mu_0^{(2)}/2$ & $\mu_0^{(2)}$ & $2 \mu_0^{(2)}$ & \\
\hline
$\sigma^{{\rm LO}}$ &
$9.61$  &  $9.65$  &  $9.63$  &  $0.41$  &  $9.65$  &  $9.71$  &  $9.71$  &  $0.62$ \\
$\Delta\sigma^{{\rm QCD}}$ &
$3.36$  &  $2.99$  &  $2.75$  &  $4.73$  &  $3.39$  &  $3.04$  &  $2.74$  &  $5.00$ \\
$\Delta\sigma^{{\rm EW}}$ &
$-0.60$ &  $-0.58$ &  $-0.55$ &  $0.39$  &  $-0.63$ &  $-0.60$ &  $-0.58$ &  $0.38$ \\
$\sigma_{q\gamma}$ &
$0.59$  &  $0.56$  &  $0.53$  &  $0.47$  &  $0.61$  &  $0.58$  &  $0.55$  &  $0.46$ \\
$\sigma_{\gamma\gamma}$ &
$0.25$  &  $0.28$  &  $0.31$  &  $0.47$  &  $0.23$  &  $0.26$  &  $0.29$  &  $0.46$ \\
$\sigma^{{\rm QCD}+{\rm EW}+q\gamma+\gamma\gamma}$ &
$13.21$ &  $12.90$ &  $12.67$ &  $4.19$  &  $13.25$ &  $12.99$ &  $12.71$ &  $4.16$ \\
\hline
\hline
\end{tabular}
\caption{Scale dependence of the LO and ${\rm QCD}+{\rm EW}+q\gamma+\gamma\gamma$ corrected integrated cross sections for the $W^-W^+H$ production at the $14~ {\rm TeV}$ LHC.}
\label{total-scale}
\end{center}
\end{table}

\par
\begin{figure}[htbp]
\begin{center}
\includegraphics[scale=0.4]{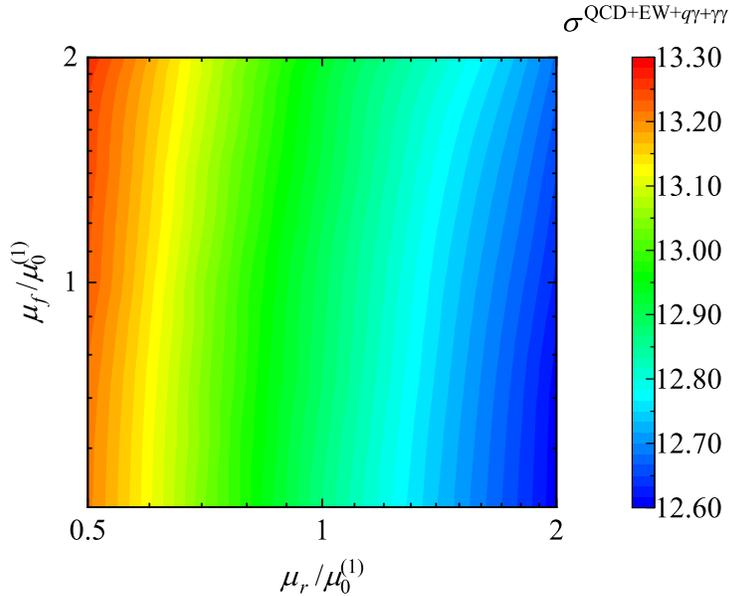}
\caption{${\rm QCD}+{\rm EW}+q\gamma+\gamma\gamma$ corrected integrated cross section for $pp \rightarrow W^-W^+H + X$ at the $14~ {\rm TeV}$ LHC as a function of $\mu_r$ and $\mu_f$.}
\label{fig4}
\end{center}
\end{figure}

The ${\rm QCD}+{\rm EW}+q\gamma+\gamma\gamma$ corrected integrated cross section for $pp \rightarrow W^-W^+H + X$ at the $14~ {\rm TeV}$ LHC as a function of $\mu_r$ and $\mu_f$, where $\mu_r$ and $\mu_f$ are two independent variables varying in the range of $\mu_r,\, \mu_f \in \big[ \mu_0^{(1)}/2,\, 2 \mu_0^{(1)} \big]$, is depicted in Fig.\ref{fig4}. From this contour plot we can draw the following conclusions: (1) The ${\rm QCD}+{\rm EW}+q\gamma+\gamma\gamma$ corrected integrated cross section increases with the increment of $\mu_f$, while decreases with the increment of $\mu_r$. (2) The $\mu_r$ dependence is much larger than the $\mu_f$ dependence. For example,
\begin{eqnarray}
&&
\sigma^{{\rm QCD}+{\rm EW}+q\gamma+\gamma\gamma}\left( x_r = 1,~ x_f = 2 \right)
-
\sigma^{{\rm QCD}+{\rm EW}+q\gamma+\gamma\gamma}\left( x_r = 1,~ x_f = 0.5 \right)
=0.07~ {\rm fb}, \nonumber \\
&&
\sigma^{{\rm QCD}+{\rm EW}+q\gamma+\gamma\gamma}\left( x_r = 2,~ x_f = 1 \right)
-
\sigma^{{\rm QCD}+{\rm EW}+q\gamma+\gamma\gamma}\left( x_r = 0.5,~ x_f = 1 \right)
=-0.61~ {\rm fb},
~~~~~~~~
\end{eqnarray}
where $x_r \equiv\mu_r\big/\mu_0^{(1)}$ and $x_f \equiv \mu_f\big/\mu_0^{(1)}$. These two conclusions are coincident with those obtained from Table \ref{total-scale}.

\par
As the factorization scale increases from $M_T/4$ to $4 M_T$, the LO cross section of $u\bar{u}$ channel increases gently first and then decreases, the LO cross section of $d\bar{d}$ channel decreases, while the LO cross sections of $c\bar{c}$, $s\bar{s}$ and $b\bar{b}$ channels increase. So the impact of the factorization scale on the total cross section is suppressed since all the channels are summed. However, there is no cancelation between these channels when considering the renormalization scale dependence. Moreover, the renormalization scale uncertainty is underestimated at the LO, because the $W^-W^+H$ production at the LO is a pure EW process. The renormalization scale uncertainty at the NLO is much larger than that at the LO. In conclusion, we may expect that the factorization scale dependence is weak compared to the renormalization scale dependence.

\par
The LO, ${\rm QCD}+{\rm EW}+q\gamma+\gamma\gamma$ corrected integrated cross sections and the corresponding QCD, EW, $q\gamma$-induced and $\gamma\gamma$-induced (relative) corrections for the $W^-W^+H$ production at the $13$, $14~ {\rm TeV}$ LHC and a $33~ {\rm TeV}$ proton-proton collider are provided in Table \ref{total-energy}. We see that the QCD and photon-induced corrections (i.e., $\Delta\sigma^{{\rm QCD}}$, $\sigma_{q\gamma}$ and $\sigma_{\gamma\gamma}$) are positive and increase as the increment of the $pp$ colliding energy, while the EW correction from the $q\bar{q}$ annihilation channel is negative and decreases as the increment of the colliding energy. The QCD correction is significant and the relative correction can reach about $37\%$ at a $33~ {\rm TeV}$ proton-proton collider. For the EW correction, the $q\gamma$-induced and $\gamma\gamma$-induced relative corrections increase quickly from $5.5\%$ to $11.6\%$ and from $2.7\%$ to $4.6\%$, respectively, while the EW relative correction from the $q\bar{q}$ annihilation channel holds steady at $-6 \sim -7\%$, as the $pp$ colliding energy increases from $13$ to $33~ {\rm TeV}$. The ratio of the full EW correction to the QCD correction, $\delta_{{\rm EW}}^{({\rm full})}\big/\delta_{{\rm QCD}}$, is about $9\%$ at the $14~ {\rm TeV}$ LHC and can exceed $25\%$ at a $33~ {\rm TeV}$ proton-proton collider. It is concluded that the photon-induced correction would be the dominant EW contribution and the full EW correction becomes more and more important with the increment of the $pp$ colliding energy.
\begin{table}[htbp]
\begin{center}
\renewcommand\arraystretch{1.8}
\begin{tabular}{ccccccc}
\hline
\hline
\multirow{2}*{$\sqrt{S}$ (TeV)}
& \multicolumn{6}{c}{Cross section~(fb)}  \\
\cline{2-7}
& $\sigma^{{\rm LO}}$ & $\Delta\sigma^{{\rm QCD}}$ & $\Delta\sigma^{{\rm EW}}$ & ~~$\sigma_{q\gamma}$~~ & ~~$\sigma_{\gamma\gamma}$~~ & $\sigma^{{\rm QCD}+{\rm EW}+q\gamma+\gamma\gamma}$ \\
\hline
$13$  &  $8.56$  &  $2.60$  &  $-0.51$  &  $0.47$  &  $0.23$  &  $11.35$  \\
$14$  &  $9.65$  &  $2.99$  &  $-0.58$  &  $0.56$  &  $0.28$  &  $12.90$  \\
$33$  &  $33.87$ &  $12.66$ &  $-2.27$  &  $3.93$  &  $1.56$  &  $49.75$  \\
\hline
\hline
\multirow{2}*{$\sqrt{S}$ (TeV)}
& & \multicolumn{4}{c}{Relative correction~($\%$)} & \\
\cline{3-6}
& & $\delta_{{\rm QCD}}$ & $\delta_{{\rm EW}}$ & $\delta_{q\gamma}$ & $\delta_{\gamma\gamma}$ & \\
\hline
$13$  &&   $30.4$  &  $-6.0$  &  $5.5$   &  $2.7$  &  \\
$14$  &&   $31.0$  &  $-6.0$  &  $5.8$   &  $2.9$  &  \\
$33$  &&   $37.4$  &  $-6.7$  &  $11.6$  &  $4.6$  &  \\
\hline
\hline
\end{tabular}
\caption{LO, ${\rm QCD}+{\rm EW}+q\gamma+\gamma\gamma$ corrected integrated cross sections and the corresponding (relative) corrections for the $W^-W^+H$ production at the $13$, $14~ {\rm TeV}$ LHC and a $33~ {\rm TeV}$ $pp$ collider.}
\label{total-energy}
\end{center}
\end{table}

\par
PDF is another source of the theoretical error for scattering processes at hadron colliders. In this work we adopt the ${\rm LUXqed\_plus\_PDF4LHC15\_nnlo\_100}$ PDFs, which contains $N = 108$ PDF sets. The PDF uncertainties of the LO and ${\rm QCD}+{\rm EW}+q\gamma+\gamma\gamma$ corrected integrated cross sections are given by \cite{Yong-Bai:2016sal,Alekhin:2011sk}
\begin{eqnarray}
\varepsilon_{{\rm PDF}}^{{\rm X}}
=
\dfrac{1}{\sigma^{{\rm X}}}
\left[
\sum_{i=1}^{N-1}
\big(
\sigma^{{\rm X}}_i
-
\sigma^{{\rm X}}_0
\big)^2
\right]^{1/2},
\end{eqnarray}
where ${\rm X} \in \big\{ {\rm LO},~ {\rm QCD}+{\rm EW}+q\gamma+\gamma\gamma \big\}$, and $\sigma_i^{{\rm X}}~ (i = 0, ..., N-1)$ are the corresponding cross sections calculated by using the $i$-th ${\rm LUXqed\_plus\_PDF4LHC15\_nnlo\_100}$ PDF set. Then we obtain
\begin{eqnarray}
\varepsilon_{{\rm PDF}}^{{\rm LO}} = 1.9\%,
~~~~~~~~~~~~
\varepsilon_{{\rm PDF}}^{{\rm QCD}+{\rm EW}+q\gamma+\gamma\gamma} = 1.7\%.
\end{eqnarray}
It clearly shows that the PDF uncertainty of the ${\rm QCD}+{\rm EW}+q\gamma+\gamma\gamma$ corrected cross section is almost the same as the PDF uncertainty of the LO cross section, and thus the PDF uncertainties from the QCD, EW, $q\gamma$-induced and $\gamma\gamma$-induced corrections are negligible. We also employ the ${\rm NNPDF23\_nlo\_as\_0119\_qed}$ PDFs in the initial-state parton convolution for comparison, and find that the PDF uncertainties obtained by using the LUXqed PDFs are much less than the corresponding ones by using the NNPDF23 PDFs. The small photon PDF uncertainty of the LUXqed PDFs was also discussed in Refs. \cite{Manohar:2016nzj,Denner:2016wet}. Compared to the scale uncertainty, the PDF uncertainty of the integrated cross section is much smaller, especially at the QCD+EW NLO. Thus, we do not consider the PDF uncertainty in estimating the theoretical error for the $W^-W^+H$ production at the LHC.

\subsection{Kinematic distributions}
\par
In this subsection, we present the LO and ${\rm QCD}+{\rm PS}+{\rm EW}+q\gamma+\gamma\gamma$ corrected kinematic distributions of final $W^{\pm}$ and Higgs bosons and their decay products for the $W^-W^+H$ production at the $14~ {\rm TeV}$ LHC.

\subsubsection{Distributions for $pp \rightarrow W^- W^+ H + X$}
\par
\begin{figure}[htbp]
\begin{center}
\includegraphics[scale=0.35]{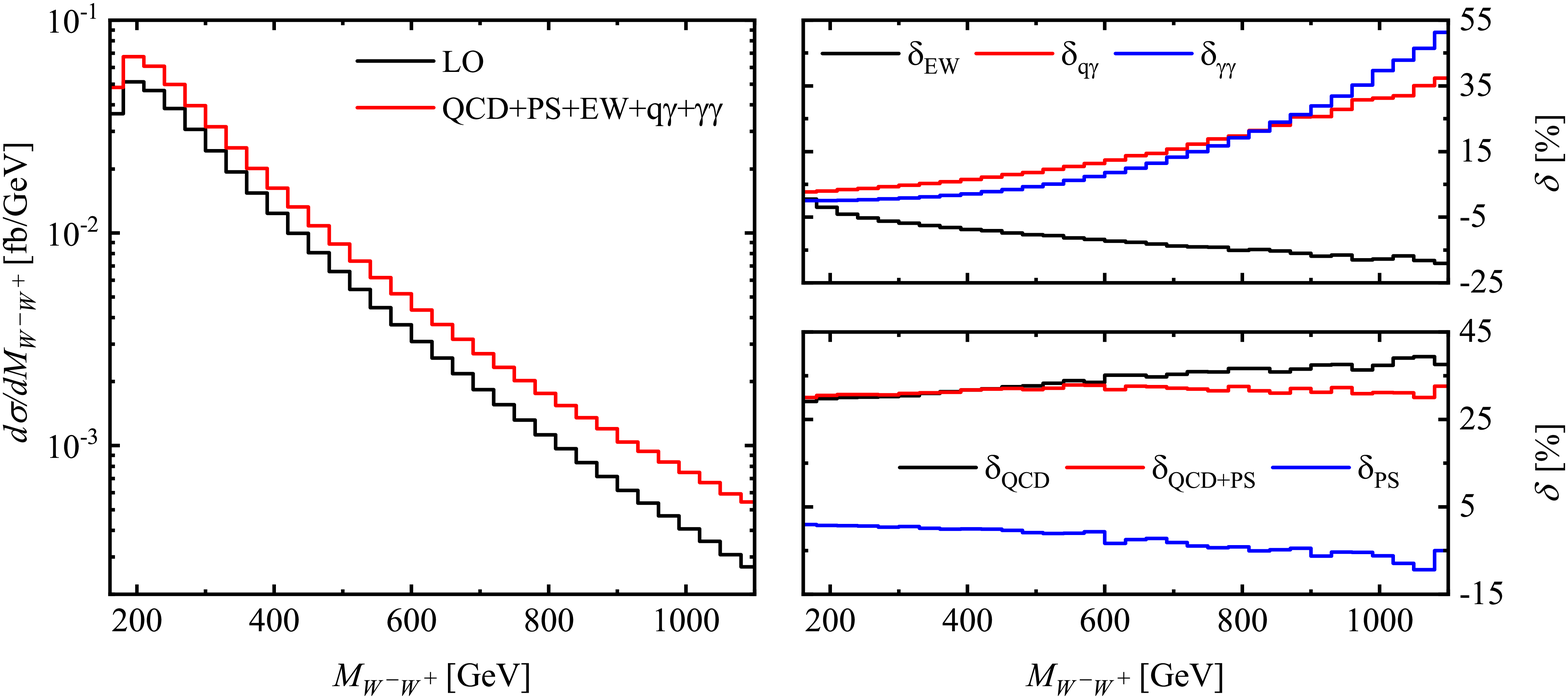}
\caption{$W$-boson pair invariant mass distributions and the corresponding relative corrections for $pp \rightarrow W^-W^+H+X$ at the $14$ TeV LHC.}
\label{fig5}
\end{center}
\end{figure}

The LO and ${\rm QCD}+{\rm PS}+{\rm EW}+q\gamma+\gamma\gamma$ corrected invariant mass distributions of the $W$-boson pair are plotted in the left panel of Fig.\ref{fig5}. The corresponding relative corrections induced by the electroweak and strong interactions ($\delta_{{\rm EW}}$, $\delta_{q\gamma}$, $\delta_{\gamma\gamma}$ and $\delta_{{\rm QCD}}$, $\delta_{{\rm QCD+PS}}$, $\delta_{{\rm PS}}$) are provided in the top right and bottom right panels, respectively\footnote{The PS relative correction is given by $\delta_{{\rm PS}} = \delta_{{\rm QCD+PS}} - \delta_{{\rm QCD}}$.}. Both the LO and ${\rm QCD}+{\rm PS}+{\rm EW}+q\gamma+\gamma\gamma$ corrected invariant mass distributions of the $W$-boson pair reach their maxima in the vicinity of $M_{W^-W^+} \sim 200~ {\rm GeV}$, and then drop down approximately logarithmically with the increment of $M_{W^-W^+}$. The QCD+PS correction enhances the LO $W$-boson pair invariant mass distribution significantly in the whole plotted $M_{W^-W^+}$ region. The corresponding QCD relative correction gradually increases from $30\%$ to approximately $40\%$ as the increment of $M_{W^-W^+}$, while the QCD+PS relative correction holds steady at about $30\%$. For the EW correction, the contribution from the $q\bar{q}$ annihilation channel suppresses the LO $W$-boson pair invariant mass distribution and the relative correction $\delta_{{\rm EW}}$ decreases from $0$ to $-20\%$, while the $q\gamma$-induced and $\gamma\gamma$-induced contributions enhance the LO $W$-boson pair invariant mass distribution and the corresponding relative corrections $\delta_{q\gamma}$ and $\delta_{\gamma\gamma}$ increase rapidly from $3\%$ to approximately $35\%$ and from $0$ to about $50\%$, respectively, as $M_{W^-W^+}$ increases from $2 M_W$ to $1.1~ {\rm TeV}$. It clearly shows that the full photon-induced correction, given by $\sigma_{\gamma{\rm -induced}} = \sigma_{q\gamma}+\sigma_{\gamma\gamma}$, is larger than the QCD+PS correction in the high $M_{W^-W^+}$ region. Thus, the LO $W$-boson pair invariant mass distribution is mainly enhanced by the QCD+PS correction in the low $M_{W^-W^+}$ region, but mainly enhanced by the photon-induced corrections in the high $M_{W^-W^+}$ region. The extremely large $\gamma\gamma$-induced correction at high invariant mass can also been seen in the invariant mass distribution of the $W^-W^+Z$ system for $pp \rightarrow W^-W^+Z + X$ at the LHC\cite{Nhung:2013jta}\footnote{We calculate the cross sections for the $\gamma \gamma \rightarrow W^-W^+H$ and $u\bar{u} \rightarrow W^-W^+H$ channels, separately. The partonic cross section for the $\gamma\gamma$ fusion channel increases, while the cross section for the $u\bar{u}$ annihilation channel decreases, as the increment of the partonic colliding energy. We also investigate the photon and $u$-quark PDFs. We find that both PDFs decrease with the increment of Bjorken $x$, and the photon PDF decreases even faster than $u$-quark PDF. Thus, the enhancement of the effects of the photon-induced contributions in the high invariant mass region mainly due to the behaviour of partonic cross section. As for other kinematic distributions such as $p_T$ distributions, since $p_T$ is different from invariant mass and it is not directly related to the partonic colliding energy, the enhancement in the high $p_T$ region is not obvious.}. The considerable negative EW correction from the $q\bar{q}$ annihilation channel in the high $M_{W^-W^+}$ region is due to the well-known Sudakov double logarithms arising from the exchange of a virtual massive gauge boson in the loops.\cite{Nhung:2013jta,Yong-Bai:2016sal}. This large Sudakov virtual correction is compensated by the positive photon-induced corrections ($\sigma_{q\gamma}$ and $\sigma_{\gamma\gamma}$) obviously, and the full EW correction enhances the LO $W$-boson pair invariant mass distribution. Therefore, the photon-induced channels should be considered for precision predictions at high energy colliders, especially in the high energy phase-space region.

\par
\begin{figure}[htbp]
\begin{center}
\includegraphics[scale=0.35]{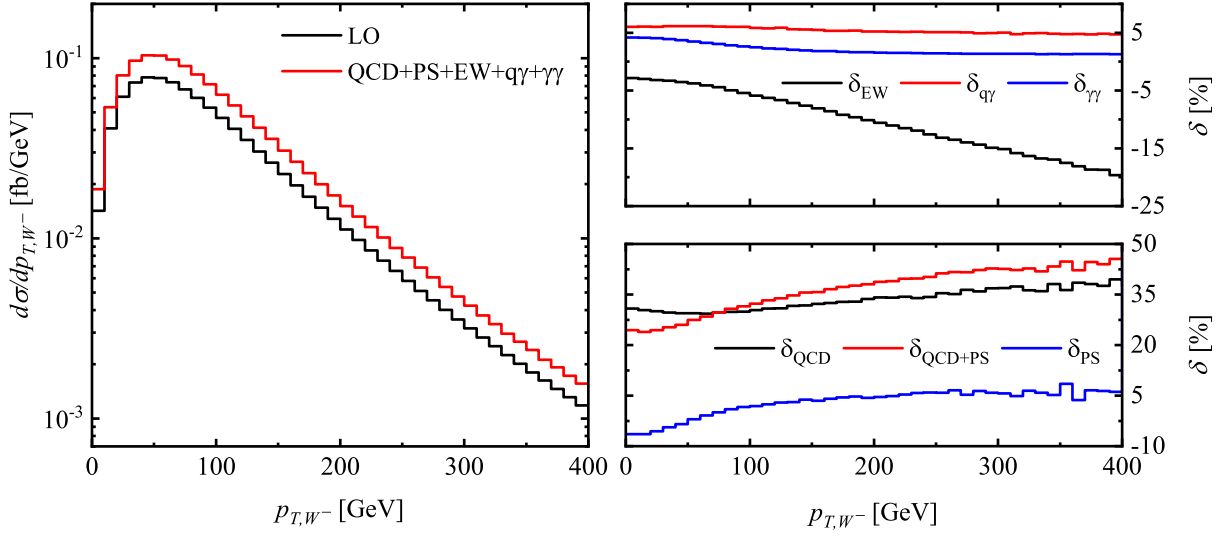}
\caption{The same as Fig.\ref{fig5}, but for the transverse momentum distribution of $W^-$.}
\label{fig6}
\end{center}
\end{figure}

The LO and ${\rm QCD}+{\rm PS}+{\rm EW}+q\gamma+\gamma\gamma$ corrected transverse momentum distributions of the $W^-$-boson and the corresponding relative corrections are presented in Fig.\ref{fig6}. Both the LO and ${\rm QCD}+{\rm PS}+{\rm EW}+q\gamma+\gamma\gamma$ corrected distributions reach their peaks at $p_{T, W^-} \sim 45~ {\rm GeV}$ and then decrease consistently as the increment of $p_{T, W^-}$. The QCD and QCD+PS relative corrections range from $30\%$ to $40\%$ and from $25\%$ to $45\%$, respectively, as $p_{T, W^-} \in [0,\, 400]~ {\rm GeV}$. As $p_{T, W^-}$ increases from $0$ to $400~ {\rm GeV}$, the EW relative correction from the $q\bar{q}$ annihilation channel decreases from about $-3\%$ to $-20\%$, while the relative corrections from the $q\gamma$ and $\gamma\gamma$ scattering channels are steady at about $5\%$ and $1 \sim 4\%$, respectively. The total production cross section is dominated by the contribution from the low $p_{T, W^-}$ region. In the vicinity of $p_{T, W^-} \sim 150~ {\rm GeV}$, the EW correction from the $q\bar{q}$ annihilation channel is almost compensated by the photon-induced corrections.

\par
\begin{figure}[htbp]
\begin{center}
\includegraphics[scale=0.35]{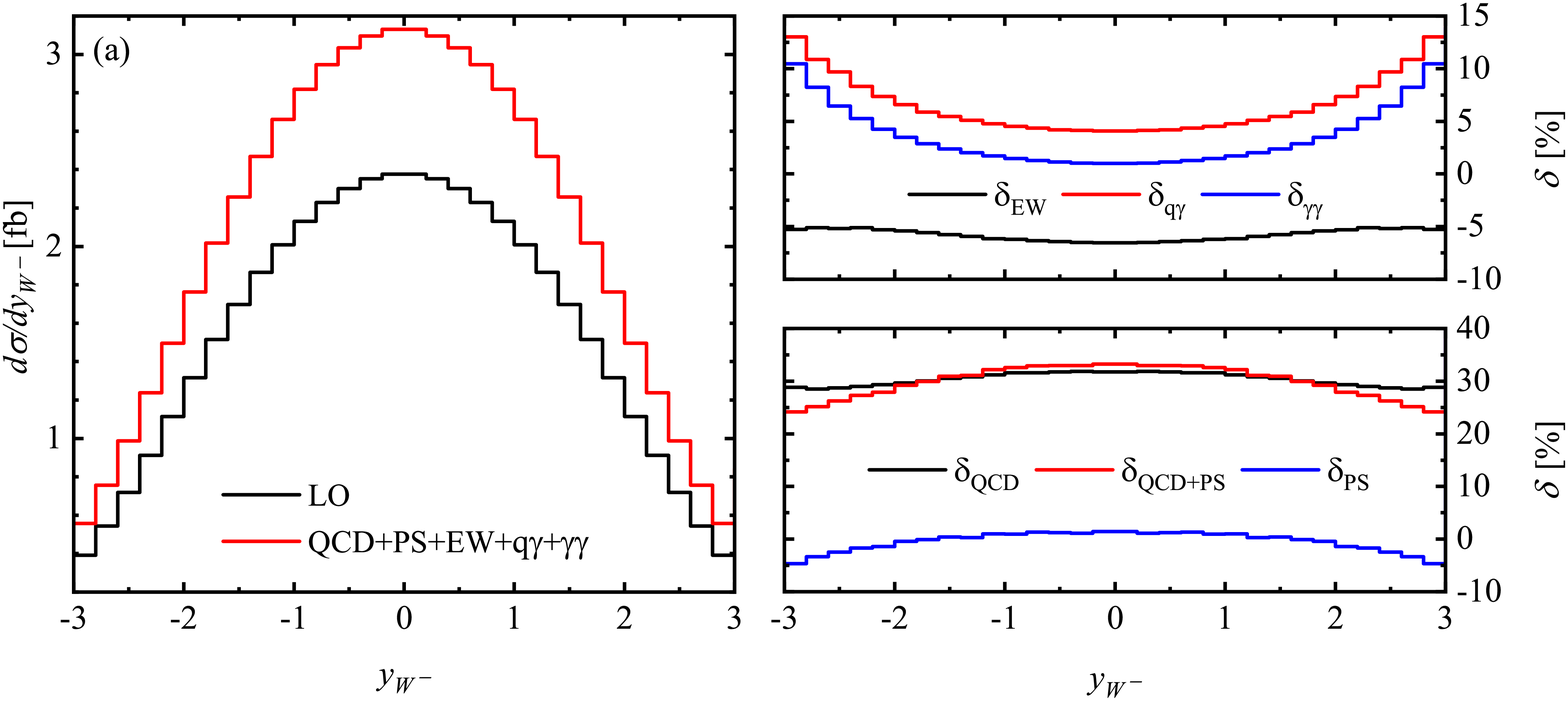}
\includegraphics[scale=0.35]{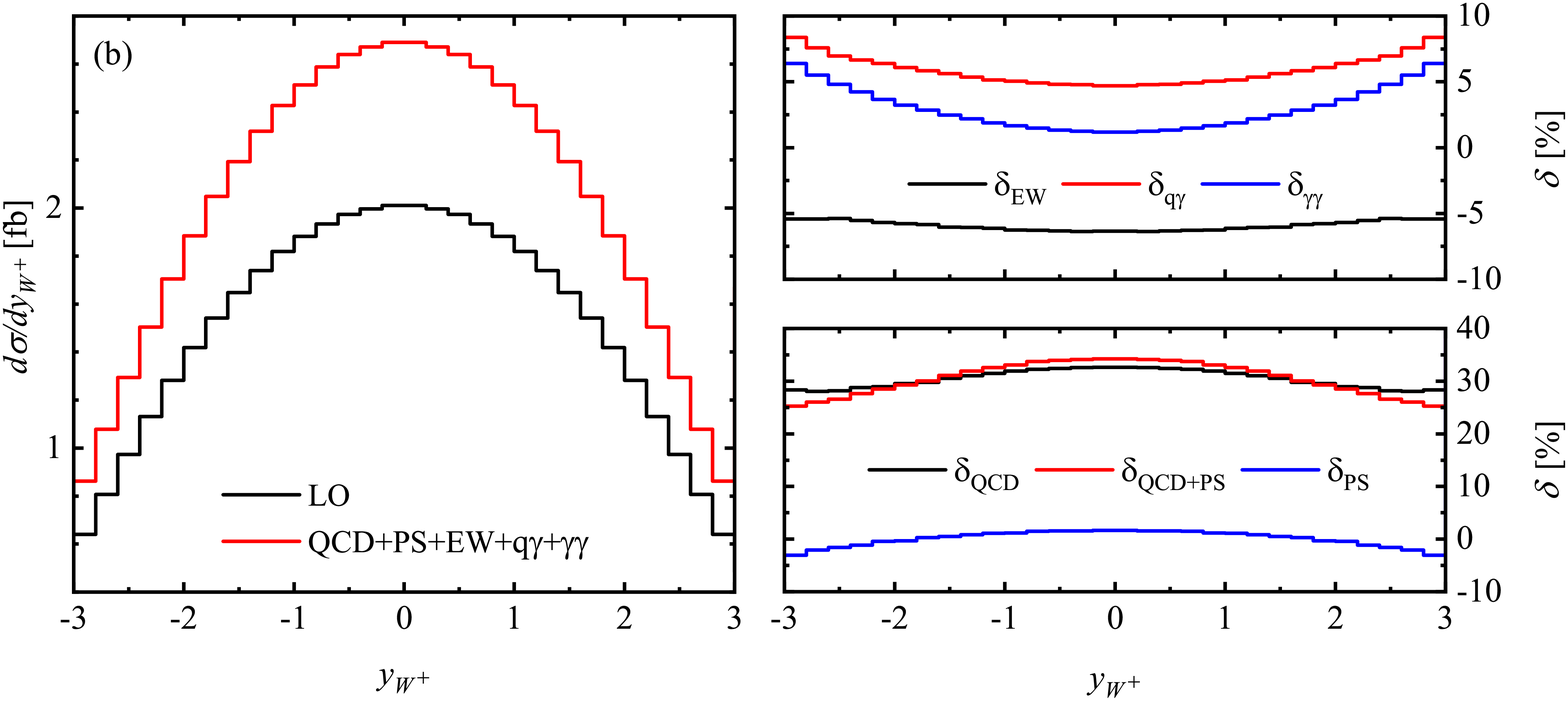}
\caption{The same as Fig.\ref{fig5}, but for the rapidity distributions of $W^{-}$ and $W^+$.}
\label{fig7}
\end{center}
\end{figure}

The LO and ${\rm QCD}+{\rm PS}+{\rm EW}+q\gamma+\gamma\gamma$ corrected rapidity distributions of $W^-$ and $W^+$ are depicted in the left panels of Figs.\ref{fig7}(a) and \ref{fig7}(b), respectively. The corresponding relative corrections are shown in the right panels. Both the LO and ${\rm QCD}+{\rm PS}+{\rm EW}+q\gamma+\gamma\gamma$ corrected rapidity distributions of $W^-$ are slightly larger than the corresponding ones of $W^+$ in the central rapidity region, but a little less than those of $W^+$ in the forward-backward rapidity region. The QCD correction enhances the LO $W$-boson rapidity distributions significantly, and the QCD+PS relative correction decreases from approximately $35\%$ to $25\%$ for both $W^-$ and $W^+$ rapidity distributions as $\vert y_W \vert$ increases from $0$ to $3$. The EW relative correction from the $q\bar{q}$ annihilation channel is negative, and insensitive to the rapidities of $W^-$ and $W^+$. It holds steady at about $-5\%$ in the whole plotted $y_W$ region. The $q\gamma$-induced relative correction increases from about $5\%$ to $13\%$ and $8\%$ for the rapidity distributions of $W^-$ and $W^+$, respectively, with the increment of $\vert y_W \vert$ from $0$ to $3$. Compared to the $q\gamma$-induced relative correction, the $\gamma\gamma$-induced relative correction is relatively small. It is less than $10\%$ and increases slowly as the increment of $\vert y_W \vert$ for both $W^-$ and $W^+$ rapidity distributions in the region of $\vert y_W \vert < 3$. The full photon-induced relative correction is sizeable, especially in the forward-backward rapidity region. We again see the importance of the $q\gamma$ and $\gamma\gamma$ scattering channels and the cancelation between the photon-induced and $q\bar{q}$-initiated EW corrections in the $W^-$ and $W^+$ rapidity distributions.

\par
\begin{figure}[htbp]
\begin{center}
\includegraphics[scale=0.35]{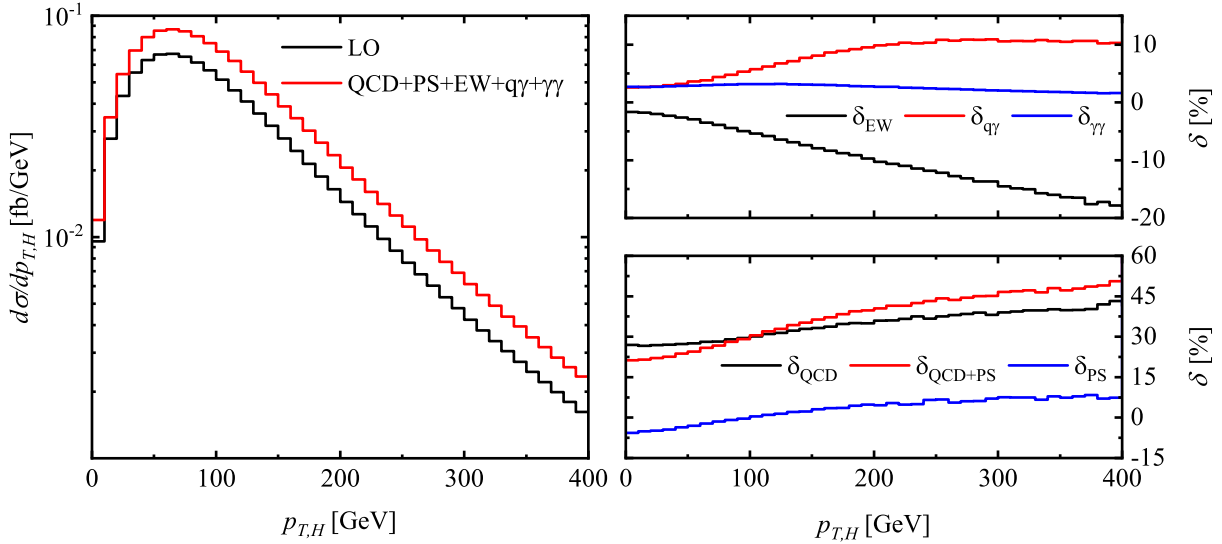}
\caption{The same as Fig.\ref{fig5}, but for the Higgs transverse momentum distribution.}
\label{fig8}
\end{center}
\end{figure}

The Higgs-boson transverse momentum distributions and the corresponding strong and electroweak relative corrections are displayed in the left and right panels of Fig.\ref{fig8}, respectively. Both the LO and ${\rm QCD}+{\rm PS}+{\rm EW}+q\gamma+\gamma\gamma$ corrected Higgs transverse momentum distributions increase sharply in the low $p_{T, H}$ region ($p_{T, H} < 50~ {\rm GeV}$), reach their maxima at $p_{T, H} \sim 65~ {\rm GeV}$, and decrease approximately logarithmically when $p_{T, H} > 80~ {\rm GeV}$ as the increment of $p_{T, H}$. The QCD+PS relative correction is positive, and increases from about $20\%$ to $50\%$ as $p_{T, H}$ increases from $0$ to $400~ {\rm GeV}$. The $q\gamma$-induced relative correction is steady at about $3\%$ in the low $p_{T, H}$ region and $10\%$ in the region of $p_{T, H} > 250~ {\rm GeV}$, respectively, and increases smoothly in the intermediate $p_{T, H}$ region ($p_{T, H} \in [50,\,  250]~ {\rm GeV}$), while the $\gamma\gamma$-induced relative correction is $2 \sim 3\%$ in the whole plotted $p_{T, H}$ region. The EW correction from the $q\bar{q}$ annihilation channel always suppresses the LO distribution, and the corresponding EW relative correction decreases from $-2\%$ to $-18\%$ as $p_{T, H}$ varies from $0$ to $400~ {\rm GeV}$. In the high $p_{T, H}$ region, the $q\bar{q}$-initiated EW correction is sizable and its absolute value is comparable to the QCD correction due to the EW Sudakov logarithms.

\par
\begin{figure}[htbp]
\begin{center}
\includegraphics[scale=0.35]{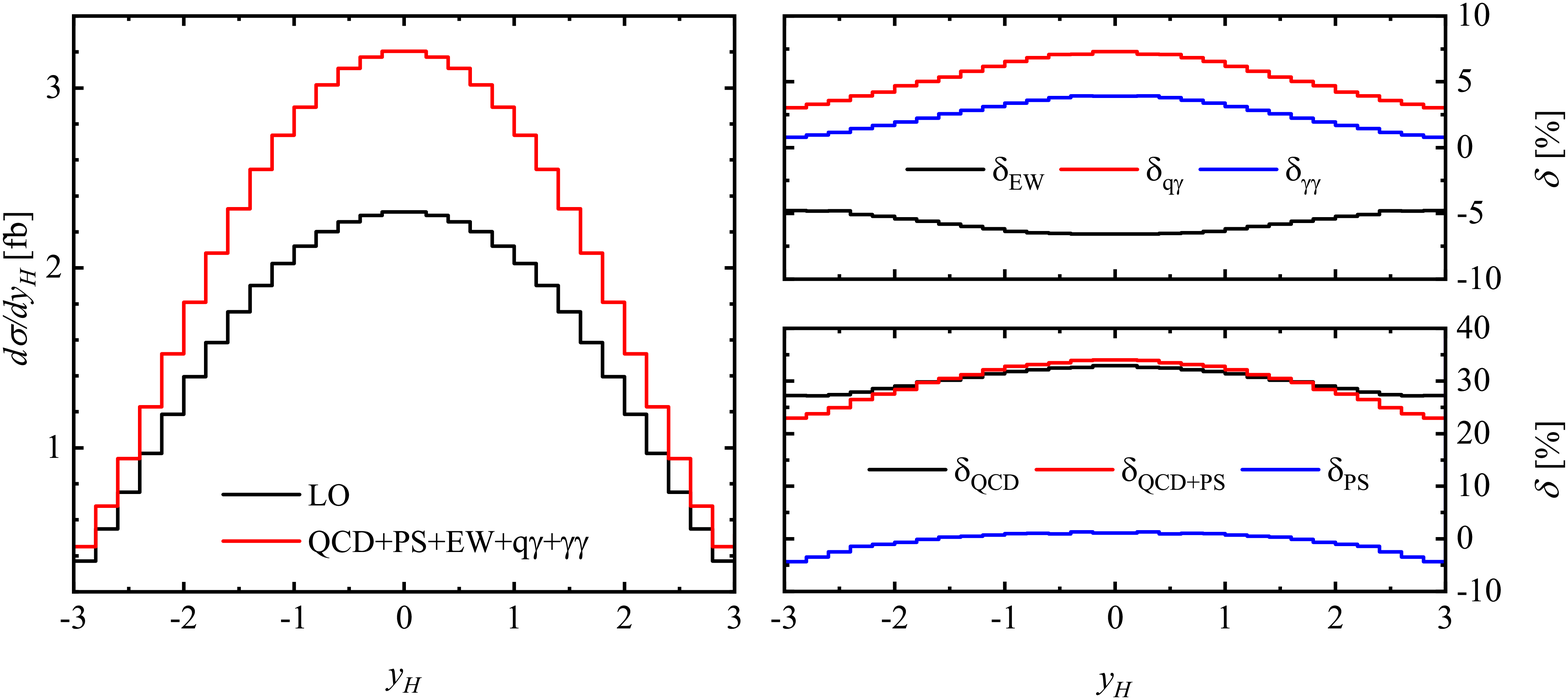}
\caption{The same as Fig.\ref{fig5}, but for the Higgs rapidity distribution.}
\label{fig9}
\end{center}
\end{figure}

The LO and ${\rm QCD}+{\rm PS}+{\rm EW}+q\gamma+\gamma\gamma$ corrected Higgs-boson rapidity distributions and the corresponding relative corrections are shown in Fig.\ref{fig9}. Analogous to the $W^-$ and $W^+$ rapidity distributions, the QCD+PS relative correction to the Higgs rapidity distribution decreases from $35\%$ to about $25\%$ as $\vert y_H \vert$ increases from $0$ to $3$, and the EW relative correction from the $q\bar{q}$ annihilation channel is insensitive to the Higgs rapidity and varies in the vicinity of $-5\%$ when $y_H \in [-3,\, 3]$. Both $q\gamma$- and $\gamma\gamma$-induced relative corrections are less than $10\%$ in the region of $\vert y_H \vert < 3$. However, it should be noted that the photon-induced relative corrections to the Higgs rapidity distribution behave quite differently from those to the $W^{\pm}$ rapidity distributions. They decrease slowly with the increment of $\vert y_H \vert$ in the plotted $y_H$ range.

\subsubsection{Distributions for $pp \rightarrow W^-W^+H + X \rightarrow l^+ l^- \nu_{l} \bar{\nu}_{l} b \bar{b} + X$}
\par
Now we turn to the $W^-W^+H$ production with subsequent $W^{\pm} \rightarrow l^{\pm} \overset{ _{(-)}}{\nu_{l}}$ and $H \rightarrow b\bar{b}$ decays at the $14~ {\rm TeV}$ LHC. The spin correlation and finite-width effects of the intermediate Higgs and $W^{\pm}$ bosons are taken into account by adopting the {\sc MadSpin} method.

\par
\begin{figure}[htbp]
\begin{center}
\includegraphics[scale=0.35]{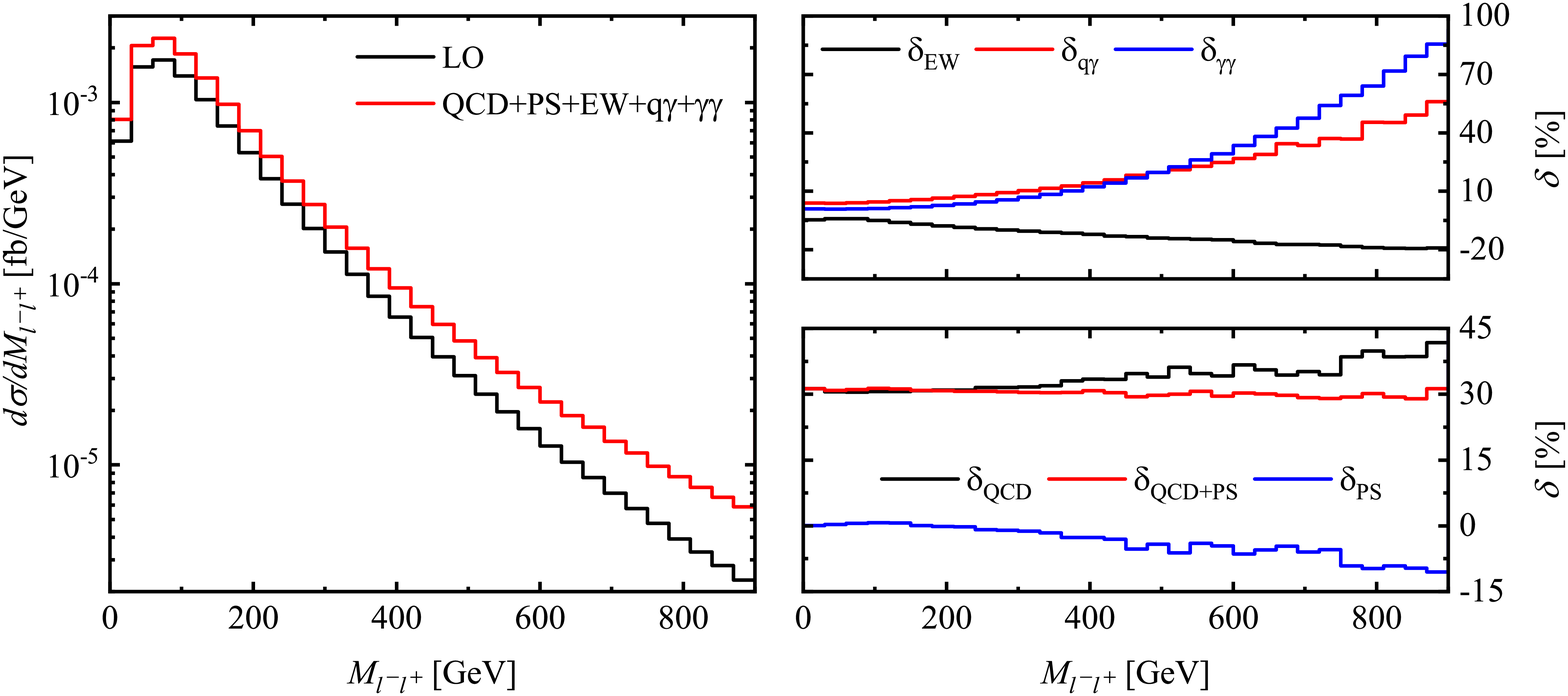}
\caption{Invariant mass distributions of the charged lepton pair and the corresponding relative corrections for $pp \rightarrow W^-W^+H \rightarrow l^+ l^- \nu_{l} \bar{\nu}_{l} b \bar{b} + X$ at the $14$ TeV LHC.}
\label{fig10}
\end{center}
\end{figure}

In the left panel of Fig.\ref{fig10}, we present the LO and ${\rm QCD}+{\rm PS}+{\rm EW}+q\gamma+\gamma\gamma$ corrected invariant mass distributions of the final charged lepton pair for $pp \rightarrow W^-W^+H + X \rightarrow l^+ l^- \nu_{l}  \bar{\nu}_{l} b \bar{b} + X$. The corresponding strong and electroweak relative corrections are plotted in the bottom right and top right panels, respectively. Since the charged leptons are the decay products of $W^{\pm}$ bosons, the charged lepton pair invariant mass distribution inherits the feature of the $W$-boson pair invariant mass distribution. Both the LO and ${\rm QCD}+{\rm PS}+{\rm EW}+q\gamma+\gamma\gamma$ corrected charged lepton pair invariant mass distributions peak at $M_{l^-l^+} \sim 75~ {\rm GeV}$, and decease approximately logarithmically as the increment of $M_{l^-l^+}$ in the range of $M_{l^-l^+} > 90~ {\rm GeV}$. In the low $M_{l^-l^+}$ region, the ${\rm QCD}+{\rm PS}+{\rm EW}+q\gamma+\gamma\gamma$ correction is dominated by the QCD contribution, while in the high $M_{l^-l^+}$ region, the LO charged lepton pair invariant mass distribution is mainly enhanced by the photon-induced corrections. For example, the relative corrections from the $q\gamma$ and $\gamma\gamma$ scattering channels can reach about $55\%$ and $85\%$, respectively, at $M_{l^-l^+} = 900~ {\rm GeV}$, while the QCD+PS relative correction is about $30\%$ in the whole plotted $M_{l^-l^+}$ region. Thus, the photon-induced channels are nonnegligible for precision measurement of $pp \rightarrow W^-W^+H + X \rightarrow l^+ l^- \nu_{l}  \bar{\nu}_{l} b \bar{b} + X$ at the $14~ {\rm TeV}$ LHC and future high energy hadron colliders, especially for large $M_{l^-l^+}$.

\par
\begin{figure}[htbp]
\begin{center}
\includegraphics[scale=0.35]{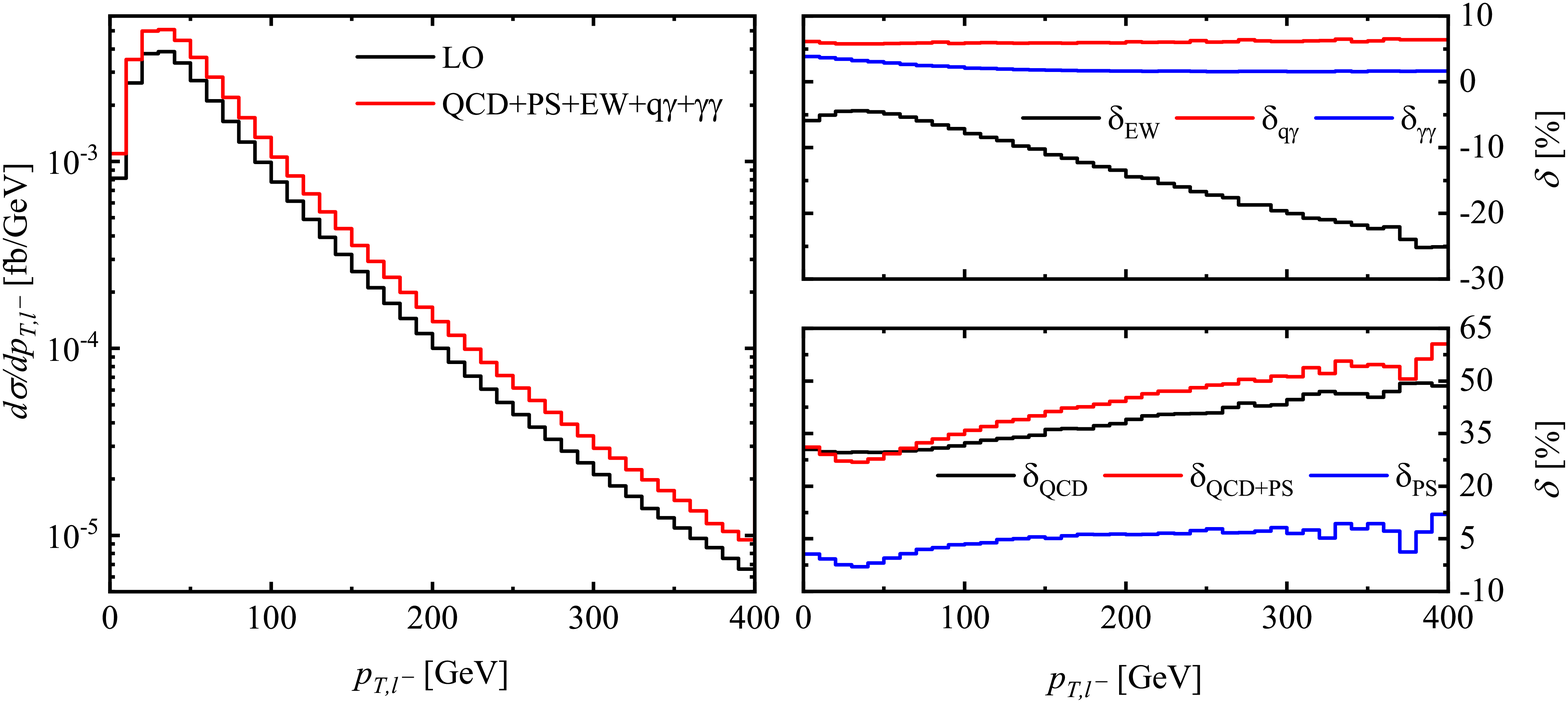}
\caption{The same as Fig.\ref{fig10}, but for the transverse momentum distribution of $l^-$.}
\label{fig11}
\end{center}
\end{figure}

Since the transverse momentum distribution of $l^+$ is similar to that of $l^-$, we only depict the transverse momentum distributions of $l^-$ and the corresponding relative corrections in Fig.\ref{fig11}. Both the LO and ${\rm QCD}+{\rm PS}+{\rm EW}+q\gamma+\gamma\gamma$ corrected $p_{T, l^-}$ distributions reach their maxima at $p_{T, l^-} \sim 35~ {\rm GeV}$ and then drop down as the increment of $p_{T, l^-}$. The QCD relative correction holds steady at about $30\%$ in the low $p_{T, l^-}$ region ($p_{T, l^-} < 50~ {\rm GeV}$) and then increases to about $50\%$ as $p_{T, l^-}$ increases to $400~ {\rm GeV}$. Due to the negative PS correction in the vicinity of $p_{T, l^-} \sim 35~ {\rm GeV}$, the QCD+PS relative correction decreases firstly, reaches its minimum at $p_{T, l^-} \sim 35~ {\rm GeV}$, and then gradually increases to about $60\%$ as $p_{T, l^-}$ increases to $400~ {\rm GeV}$. The EW relative correction from the $q\bar{q}$ annihilation channel decreases consistently as the increment of $p_{T, l^-}$ in the region of $p_{T, l^-} > 50~ {\rm GeV}$, and reaches about $-25\%$ at $p_{T, l^-} = 400~ {\rm GeV}$ due to the large EW Sudakov effect. The photon-induced relative corrections are insensitive to the transverse momentum of $l^-$: The $q\gamma$-initiated relative correction holds steady at about $6\%$ and the $\gamma\gamma$-initiated relative correction is about $2 \sim 4\%$ in the whole plotted $p_{T, l^-}$ region.

\par
\begin{figure}[htbp]
\begin{center}
\includegraphics[scale=0.35]{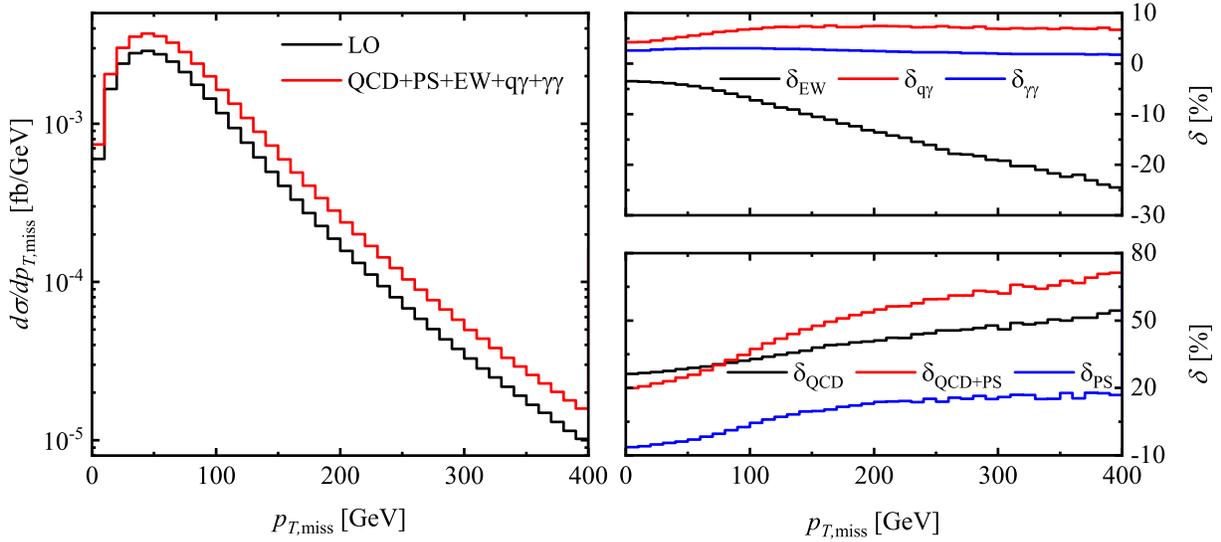}
\caption{The same as Fig.\ref{fig10}, but for the missing transverse momentum distribution.}
\label{fig12}
\end{center}
\end{figure}

The LO, ${\rm QCD}+{\rm PS}+{\rm EW}+q\gamma+\gamma\gamma$ corrected missing transverse momentum distributions and the corresponding relative corrections for $pp \rightarrow W^-W^+H + X \rightarrow l^+ l^- \nu_{l}  \bar{\nu}_{l} b \bar{b} + X$ at the $14~ {\rm TeV}$ LHC are shown in Fig.\ref{fig12}. The corrections do not distort the line shape of the LO $p_{T, {\rm miss}}$ distribution, and both the LO and ${\rm QCD}+{\rm PS}+{\rm EW}+q\gamma+\gamma\gamma$ corrected $p_{T, {\rm miss}}$ distributions reach their maxima at $p_{T, {\rm miss}} \sim 45~ {\rm GeV}$. As the increment of $p_{T, {\rm miss}}$ from $0$ to $400~ {\rm GeV}$, the QCD and QCD+PS relative corrections increase from $25\%$ to $55\%$ and from $20\%$ to $70\%$, respectively, while the EW relative correction from the $q\bar{q}$ annihilation channel decreases from $-3\%$ to $-25\%$. The relative corrections from the $q\gamma$ and $\gamma\gamma$ scattering channels are significant, but much less than the QCD+PS relative correction. They are almost independent of the missing transverse momentum, especially in the high $p_{T, {\rm miss}}$ region: The $q\gamma$-induced relative correction is about $7\%$ and the $\gamma\gamma$-induced relative correction is $2 \sim 3\%$, respectively, as $p_{T, {\rm miss}} \in [100,\, 400]~ {\rm GeV}$.

\par
\begin{figure}[htbp]
\begin{center}
\includegraphics[scale=0.35]{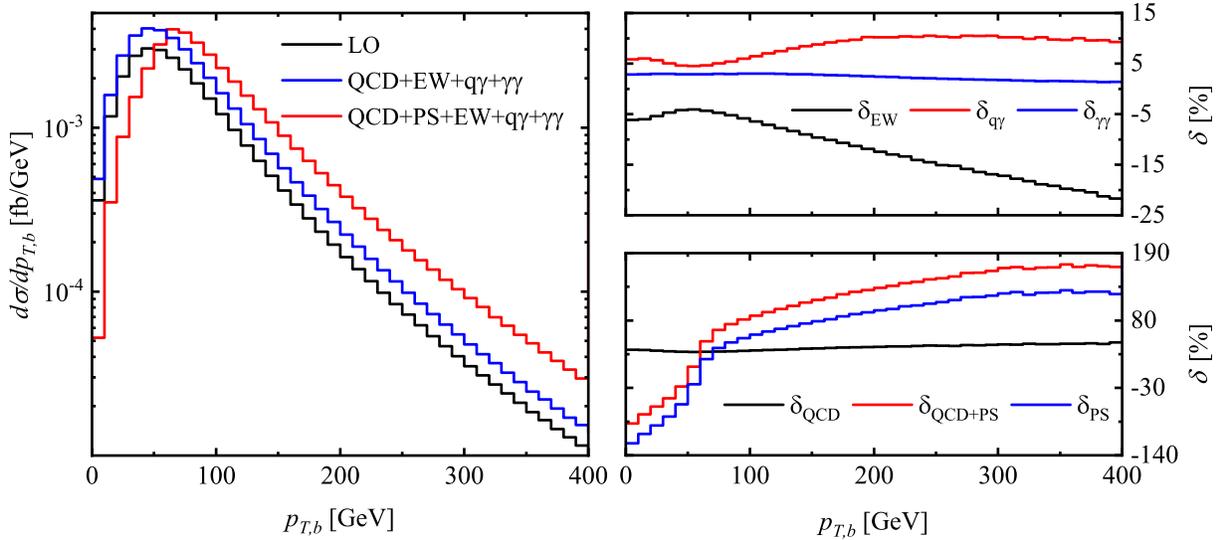}
\caption{The same as Fig.\ref{fig10}, but for the $b$-jet transverse momentum distribution.}
\label{fig13}
\end{center}
\end{figure}

The transverse momentum distribution of the final anti-bottom jet should be the same as that of the bottom jet due to the CP conservation in the $H \rightarrow b \bar{b}$ decay, so that we only study the $b$-jet transverse momentum distribution and discuss the influences of the QCD, PS, EW, $q\gamma$-induced and $\gamma\gamma$-induced corrections on the $p_{T, b}$ distribution in the following. From Fig.\ref{fig13} we see that both the LO and ${\rm QCD}+{\rm EW}+q\gamma+\gamma\gamma$ corrected $b$-jet transverse momentum distributions peak at $p_{T, b} \sim 45~{\rm GeV}$, while the ${\rm QCD}+{\rm PS}+{\rm EW}+q\gamma+\gamma\gamma$ corrected $b$-jet transverse momentum distribution reaches its maximum at $p_{T, b} \sim 65~{\rm GeV}$. Analogous to the $p_{T, W^-}$, $p_{T, H}$, $p_{T, l^-}$ and $p_{T, {\rm miss}}$ distributions, the $b$-jet transverse momentum distribution increases sharply in the low $p_{T, b}$ region and decreases approximately logarithmically after reaching its maximum as the increment of $p_{T, b}$. The QCD relative correction varies in the range of $30 \sim 45\%$, while the QCD+PS relative correction increases from approximately $-90\%$ to the order of $165\%$, as $p_{T,b}$ increases from $0$ to $400~ {\rm GeV}$. It clearly shows that the PS correction to the $b$-jet transverse momentum distribution is more significant compared to the PS corrections to the kinematic distributions of colorless particles discussed above. Generally speaking, the distributions of colored particles are more sensitive to PS effects than those of colorless particles, since the soft gluon can radiate from not only the initial-state partons but also the final-state colored particles. Thus the jet distribution are very sensitive to PS effects. The EW relative correction from the $q\bar{q}$ annihilation channel varies in the vicinity of $-5\%$ in the low $p_{T, b}$ region and then gradually decreases to approximately $-20\%$ as $p_{T, b}$ increases to $400~ {\rm GeV}$. The $q\gamma$-induced relative correction ranges from about $5\%$ to $10\%$ for $p_{T, b} < 200~ {\rm GeV}$, and holds steady at about $10\%$ when $p_{T, b} > 200~ {\rm GeV}$. The $\gamma\gamma$-induced relative correction is insensitive to the $b$-jet transverse momentum. It is less than $3\%$ and decreases very slowly as the increment of $p_{T, b}$ in the plotted $p_{T, b}$ region.

\vskip 5mm
\section{SUMMARY}
\par
The production of $W^-W^+H$ at the LHC can help to understand the EW symmetry breaking and to search for new physics beyond the SM. In this work, we calculate the shower-matched NLO QCD correction and the $q\bar{q}$-, $q\gamma$- and $\gamma\gamma$-initiated EW corrections to the $W^-W^+H + X$ production at the $14~ {\rm TeV}$ LHC. We employ four different subtraction schemes to subtract the top-resonance effect for comparison, and adopt the {\sc MadSpin} method to deal with the subsequent $W^{\pm} \rightarrow l^{\pm} \overset{ _{(-)}}{\nu_{l}}$ and $H \rightarrow b\bar{b}$ decays in order to preserve the spin correlation and finite-width effects as far as possible. The integrated cross section and some kinematic distributions of $W^{\pm}$ and $H$ and their decay products are provided. The scale and PDF uncertainties of the integrated cross section are also given for estimating the theoretical error. Our numerical results show that the theoretical error of the ${\rm QCD}+{\rm EW}+q\gamma+\gamma\gamma$ corrected integrated cross section mainly comes from the renormalization scale dependence of the QCD correction. The QCD correction enhances the LO differential cross section significantly, especially in the central rapidity and high energy regions, while the EW correction from the $q\bar{q}$ annihilation channel suppresses the LO differential cross section obviously. The QCD and $q\bar{q}$-initiated EW relative corrections to the integrated cross section are about $31\%$ and $-6\%$, respectively. The relative corrections from the photon-induced channels, $q\gamma \rightarrow W^-W^+H q$ and $\gamma\gamma \rightarrow W^-W^+H$, are insensitive to the transverse momenta of final products. The $q\gamma$- and $\gamma\gamma$-induced relative corrections to the integrated cross section are about $6\%$ and $3\%$, respectively, and can compensate the negative EW relative correction from the $q\bar{q}$ annihilation channel. The PS relative corrections to the kinematic distributions of colorless particles are ${\cal O}(10\%)$ in the bulk of the phase space, while the PS relative correction to the $b$-jet transverse momentum distribution can exceed $100\%$ in the high $p_{T, b}$ region. Thus, we should take into account the shower-matched NLO QCD correction and the EW corrections from the $q\bar{q}$ annihilation and photon-induced channels in precision study of the $W^-W^+H$ production at the LHC and future hadron colliders.

\vskip 10mm
\par
\noindent{\large\bf ACKNOWLEDGMENTS} \\
This work is supported in part by the National Natural Science Foundation of China (Grants No. 11775211 and No. 11535002), the Applied Basic Research Programs of Yunnan Provincial Science and Technology Department (Grant No. 2016FB008) and the CAS Center for Excellence in Particle Physics (CCEPP).

\end{document}